\documentclass[12pt,showpacs,preprintnumbers,amsmath,amssymb]{revtex4}
\usepackage[english]{babel}
\usepackage[latin2]{inputenc}
\usepackage{epsfig}

\begin{document}

\title{Disorder--Induced Effects in III--V Semiconductors with Mn}

\author{J.~Ma\v{s}ek and F.~M\'{a}ca}

\address{Institute of Physics, AS CR, Na Slovance 2,
CZ-182 21 Prague 8}


\begin{abstract}
The substitution of Mn in the III-V diluted magnetic
semiconductors leads to a strong electron scattering on
impurities. Besides the features induced in the valence band by
the hybridization with the Mn d--states, also the conduction band
is affected by the absence of the Mn s--states at its edge. Also
the high concentration of compensating donors modifies the band
structure. This is shown on the absorption coefficient
$\varepsilon_{2}(\omega)$ of GaP doped with Mn and Se. The
absorption evaluated by {\sl ab initio} density functional
calculations starts with a smooth tail and does not show the
structure typical for III-V materials. We analyze these features
and the role of the donors on model systems using the
tight--binding coherent potential approach.
\end{abstract}


\maketitle

\section{Introduction}

The properties of the diluted magnetic semiconductors (DMS), such
as II-VI and III-V materials doped with Mn,  are mostly determined
by the presence of open d-shells of Mn. The d-electrons form the
local magnetic moments, and the hybridization of the d-states with
the band states leads to strong magnetooptical and magnetoelectric
phenomena \cite{Furdyna88, Ohno99}. In addition, the magnetic
polarization of  the band states of the host semiconductor results
in kinetic exchange interaction between the moments.

The change of the crystal potential due to the Mn substitution has
also an important direct effect on the band states. In the
simplest description, the substitution is represented by replacing
the atomic s- and p-levels $E_{s}$ and $E_{p}$ of the host atom by
the atomic levels of the impurity. In traditional II-VI DMS with
Mn, the differences in $E_{s}$ are of order of 1 - 2 eV
\cite{tables}. This, with respect to the typical band structure of
the II-VI semiconductors, represents a perturbation of an
intermediate strength \cite{Larson88}, which can not be treated
within the virtual crystal approximation (VCA).

In the III-V semiconductors,  however, the atomic levels of Mn
differ from the atomic levels of the group-III cations even more.
The change of $E_{s}$ in GaAs and GaP doped with Mn is
approximately 4 eV \cite{tables} and also the difference of the
p-levels is remarkable. Strong impurity potential, comparable to
the bandwidth, is expected to cause a strong reconstruction of the
electronic structure of the mixed crystal. This prediction of the
tight-binding theory has been recently confirmed also by our
recent calculations \cite{Masek01, Maca02}.

We expect a particularly strong effect of the Mn substitution in
the lowest conduction bands, because it is mostly composed of the
cationic s--states. This means that the impurity scattering should
have a pronounced implication on the near-edge absorption due to
the violation of the momentum conservation and the appearance of
the non-direct transitions \cite{Spicer67} from the valence to the
conduction band.

The situation is, however, more complex. It is well established
that Mn doped III-V DMS reveal an almost complete compensation
\cite{Beschoten99}. This means that the concentration of either
intentional donors or native compensating defects increases
proportionally to the concentration of Mn \cite{Maca02}. These
donors represent an additional source of the disorder scattering.

We present here the first ab-initio study of the interband
absorption in the III-V DMS. It is combined with tight-binding CPA
calculations suitable for the interpretation of the
disorder-induced features in the electronic structure of the
dilute mixed crystals.

We concentrate on the interband absorption. That is why we
consider only the limiting case of a complete compensation, where
the IR transition inside the valence band are absent.

\section{Absorption coefficient}

Although the main interest is paid to (Ga,Mn)As, we applied our
density functional study to (Ga,Mn)P with a wider bandgap that is
less sensitive to narrowing of the band gap due to the local
density approximation. We take Se$_{\rm P}$ as the compensating
donor. The mixed crystal is represented by  a periodic system with
a large unit cell (LUC) containing 16 atoms. One molecular unit
(Ga--P) of the LUC is replaced by a close Mn--Se pair. The
electronic structure and the imaginary part of the dielectric
function, $\varepsilon_{2}(\omega)$, were calculated by the
full-potential linearized augmented-plane-waves (FPLAPW) method
\cite{WIEN97}. The spin--orbit coupling was not included in our
spin--polarized calculations. The generalized gradient
approximation to the exchange--correlation potential was used. The
results are summarized in Figs. 1 and 2.

Fig. 1 shows the spin-polarized total density of states (DOS) of
the Ga$_{7}$MnP$_{7}$Se crystal in the low--temperature,
ferromagnetic state altogether with the DOS of pure GaP. The bands
of Ga$_{7}$MnP$_{7}$Se are strongly magnetically polarized due to
high concentration (12.5 at. \%) of magnetic impurities. The spin
splitting of the valence band (1.2 eV) corresponds to the exchange
constant $J_{pd} \approx 2$ eV. The bandgaps for both majority and
minority spin electrons, 0.57 eV and 1.29 eV, respectively, are
smaller than the bandgap obtained for the pure GaP (1.67 eV).
%
In contrast to (Ga,Mn)As \cite{Maca02}, the present material with
a small overlap (0.16 eV) of both bandgaps is a semiconductor. The
width of the common bandgap is expected to increase for lower Mn
doping.

\begin{figure}[tbp]
\begin{center}
\epsfig{file=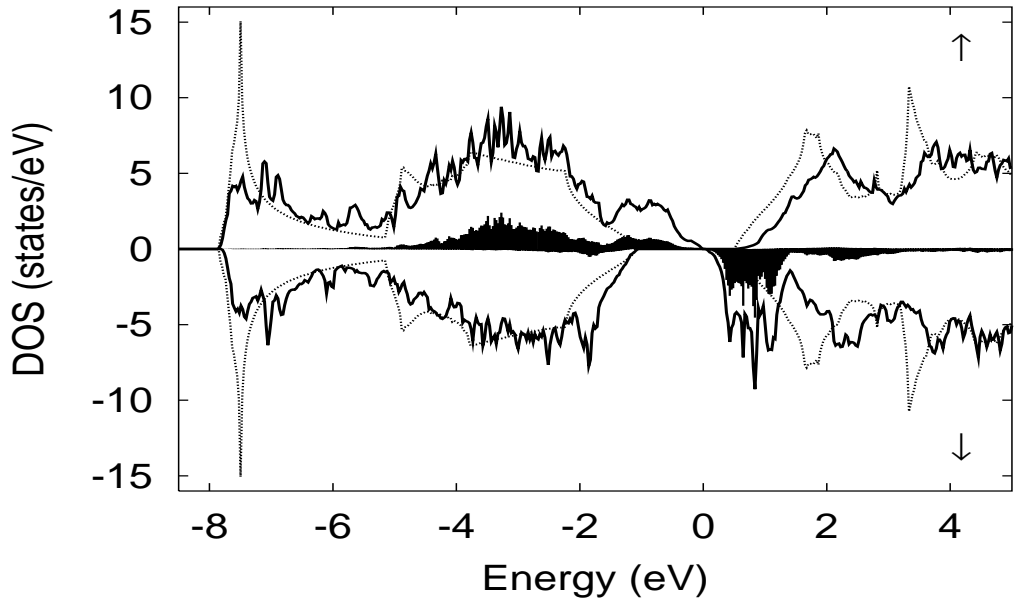, width=84mm, height=146mm, angle=270}
\end{center}
\caption{Spin polarized density of states for Ga$_{7}$MnP$_{7}$Se
(full line) compared to the rescaled DOS for pure GaP crystal
(thin line). The hatched areas show the Mn d-states.}
\end{figure}

Although the uppermost part of the majority spin DOS at the top of
the valence band can be viewed as an impurity band formed by the
acceptor states, no gap states are created due to Se impurities.
They act as donors with zero activation energy. The presence of Se
has no important effect, neither on the distribution of the Mn
d-states (see Fig. 1) nor on the total DOS around the bandgap. It
means that most of the changes in the absorption should be
attributed to Mn. The Mn s-states contribute mostly to the broad
maximum of the conduction band DOS around 4.5 eV and they are
almost completely absent at the edge of the conduction band, so
that its bottom is locally eroded around Mn. Also the empty Mn
d-states substantially modify the nature of the bottom of the
conduction band for the minority-spin electrons.

\begin{figure}[btp]
\begin{center}
\epsfig{file=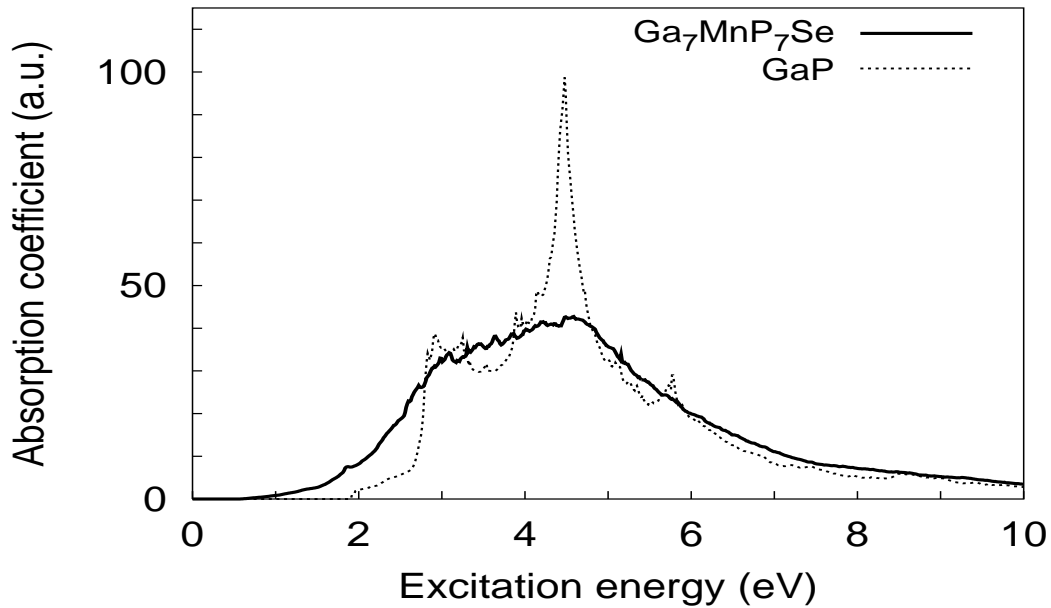, width=84mm, height=146mm, angle=270}
\end{center}
\caption{Absorption coefficient, $\varepsilon_{2}(\omega)$, for
Ga$_{7}$MnP$_{7}$Se (full line) compared to
$\varepsilon_{2}(\omega)$ for pure GaP crystal (dotted line)}
\end{figure}

In Fig. 2 we compare the calculated $\varepsilon_{2}(\omega)$ for
Ga$_{7}$MnP$_{7}$Se with the results for the GaP crystal from
near-edge to the UV range. The absorption edge of pure GaP is
followed by a steep increase of the absorption at 2.85 eV (E$_{1}$
transition) and by a main sharp peak at 4.4 eV (E$_{2}$
transition).

The change due to the Mn (and Se) doping is remarkable. The onset
of the absorption is shifted to lower energies by $\approx 1.1$
eV. Instead of well developed absorption edge, a smooth increase
of the absorption reminds an Urbach tail \cite{Urbach53}. The
interband absorption does not show any structure typical for the
zinc-blende semiconductors, only a broadened E$_{2}$ peak
survives. The E$_{1}$ transition, more sensitive to the
arrangement of the bands along the $\Lambda$ line in the Brillouin
zone, disappears completely and the absorption gradually increases
between 1.5 eV and 4 eV eV.

\section{Spectral density}

To describe the DMS with lower impurity concentrations we use the
coherent potential description (CPA) \cite{Velicky68} of the
electronic structure of the DMS. The coherent potential method,
resulting in configurationally averaged quantities, can also
easily handle the room-temperature, magnetically disordered phase
of the DMS. The avaraging over all random distributions of the
impurities restores the full translational symmetry of the
lattice, so that the wavevector $k$ can be used as a quantum
number. The spectral density $A(k,E)$, i.e. the DOS decomposed
into contributions from the various points of the Brillouin zone,
represents a detailed information about the dispersion of the
electron states.

\begin{figure}[btp]
\begin{center}
\epsfig{file=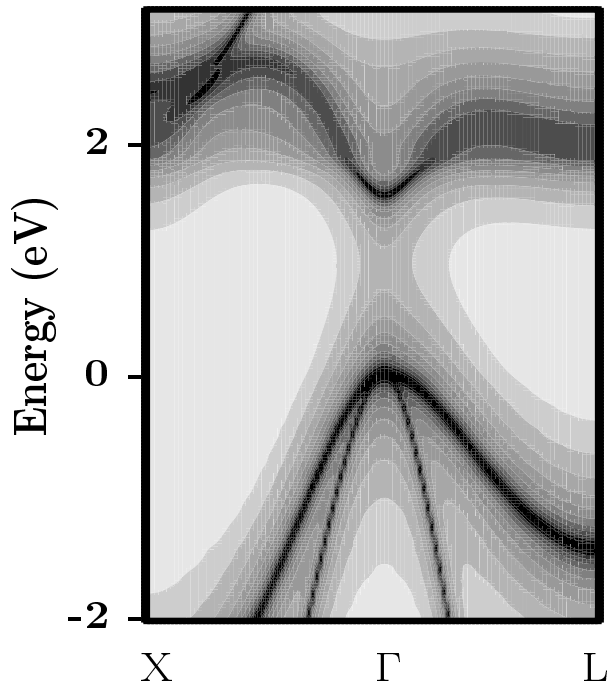, width=83mm, height=96mm}
\end{center}
\caption{Contour map of the spectral density $A(k,E)$ for the
minority spin electrons in
Ga$_{0.96}$Mn$_{0.04}$P$_{0.96}$Se$_{0.04}$ along the path
X--$\Gamma$--L in the Brillouin zone.}
\end{figure}

We apply a tight-binding version of the CPA \cite{Masek87} to GaAs
co-doped with 4 at. \% of Mn and 4 at. \% of Se. The
parametrization of the tight-binding Hamiltonian provides a
correct band gap for the pure GaAs crystal \cite{Talwar82}, as
well as appropriate exchange splitting of the Mn d--states. As
shown in Fig. 3 for minority spin electrons, the influence of the
dopants on the spectral density is strong. Similar results were
obtained for the majority spin electrons and also for the
paramagnetic phase of Ga$_{0.92}$Mn$_{0.04}$Se$_{0.04}$.

The linewidth of the Bloch states in the highest valence band
 and in the lowest conduction band is, in all cases, approximately
0.5 eV. The linewidth of the corresponding excitations, in
particular of the E$_{1}$ transitition is $\approx 1.0$ eV, large
enough to smear out the E$_{1}$ peak in $\varepsilon_{2}(\omega)$.
However, it is important to note that the effect of the impurity
broadening depends on the orbital composition of the bands and on
the position in the Brillouin zone, as seen in Fig. 3. The narrow
peaks of $A(k,E)$ indicate those parts of the bands where the
effect of the impurities is rather weak.

\section{Comparison with other donors}

The almost complete compensation \cite{Beschoten99} in the III-V
DMS with Mn means that the total concentration of donors, either
intentional or native defects, is almost equal to the
concentration of Mn, i.e. a few atomic percent. In such case, we
have mixed rather than doped semiconductors and the modification
of their band structure due to the donors must be considered. We
compare the electronic structure of GaAs with 4 at. \% of Mn and
with a corresponding concentration of the compensating donors. We
consider donors in both anion and cation sublattices (Se, Sn), and
As antisite defects.

The most favorite donors are the As antisite defects. They are
double donors and create an impurity band deep in the bandgap,
centered at 0.8 eV above the valence band. Also the co-doping with
tin results in an impurity band around 1.7 eV, which is, however,
resonant with the bottom of the conduction band. Both impurities
give an inhomogeneous broadening ($\approx 0.7$ eV) of the
conduction--band states.

Finally, the intentional compensation with Se, considered in
Sections 2 a 3, can be viewed as an alloying of the III-V
materials with MnSe, without formation of the impurity states.

\section{Summary}

The substitution of Mn in the GaAs crystal is connected with a
scattering of electrons on the dopants. It results in broadening
of the spectral lines and in smearing out the electron spectra.
This is particularly seen in the calculated absorption
coefficient. We found that the absorption edge of the compensated
(Ga,Mn)P co--doped with Se is shifted by $\approx$ 1.0 eV with
respect to pure GaP crystal.  The absorption is smooth, without
any structure typical to the E$_{1}$ transitions in the III-V
materials.

Three aspects of the electron scattering on impurities can be
distiguished: (i) Hybridization of the host states with the Mn
d--states, acting locally on the p--states of As atoms bonded to
Mn. This effect is particularly important in the upper part of the
valence band. (ii) Impurity potential acting on the cationic s-
and- p--states that modifies mostly the conduction band states.
(iii) Scattering on compensating donors that are present in real
III-V materials in a concentration comparable to the concentration
of Mn.

While (i) and (ii) are inherent to the Mn impurity, (iii) may be
optimized by the choice of the compensating donor. We found that
the effect of Se is weak in comparison to the influence of Mn. The
modification of the band states by both As antisite defects and
substitutional Sn is much stronger. These two impurities also form
impurity bands in (Ga,Mn)As while Se does not.

Assuming that the concentration of Mn determines the total number
of the donors we deduce that the presence of the intentional
donors can reduce the number of native antisite defects. If it is
so, the co-doping of the III-V DMS with Se--like impurities will
be quite important.
\\

\noindent {\bf Acknowledgment} The financial support was provided
by the Academy of Sciences of the Czech Republic (Grant No.
A1010214) and by RTN project "Computational Magnetoelectronics" of
the European Commission (HPRN-CT-2000-00143).


\end{document}